\documentclass[reprint, amsmath,amssymb, aps, superscriptaddress]{revtex4-1} 

\usepackage{graphicx}% Include figure files
\usepackage{dcolumn}% Align table columns on decimal point
\usepackage{bm}% bold math
\usepackage{color}

\begin{document}

\preprint{IPMU14-0086}

\title{Approaching conformal window of $O(n)\times O(m)$ symmetric Landau-Ginzburg models from conformal bootstrap}


\author{Yu Nakayama}
\author{Tomoki Ohtsuki}
\affiliation{Kavli Institute for the Physics and Mathematics of the Universe (WPI),  \\ Todai Institutes for Advanced Study,
University of Tokyo, \\ 
5-1-5 Kashiwanoha, Kashiwa, Chiba 277-8583, Japan} 

\date{\today}% It is always \today, today,
             %  but any date may be explicitly\input{../../../../Volumes/HD-PSU2/file_reserved/iwasaki_scgt}

\begin{abstract}
$O(n) \times O(m)$ symmetric Landau-Ginzburg models in $d=3$ dimension possess a rich structure of the renormalization group and its understanding offers a theoretical prediction of the phase diagram in frustrated spin models with non-collinear order. Depending on $n$ and $m$, they may show chiral/anti-chiral/Heisenberg/Gaussian fixed points within the same universality class. We approach all the fixed points in the conformal bootstrap program by examining the bound on the conformal dimensions for scalar operators as well as non-conserved current operators with consistency crosschecks. For large $n/m$, we show strong evidence for the existence of four fixed points by comparing the operator spectrum obtained from the conformal bootstrap program with that from the large $n/m$ analysis. We propose a novel non-perturbative approach to the determination of the conformal window in these models based on the conformal bootstrap program. From our numerical results, we predict that for $m=3$, $n=7\sim 8$ is the edge of the conformal window for the anti-chiral fixed points.
\end{abstract}

\pacs{11.25.Hf,64.60.De,64.60.fd}% PACS, the Physics and Astronomy
\maketitle 
\section{Introduction} 
Conformal field theories (CFTs) have played central roles in theoretical physics since they lie at the endpoints of generic renormalization group (RG) flows. In the study of critical phenomena, it is crucial to understand what kind of fixed points are present with the desired symmetries (see e.g. \cite{Cardy:1996xt}\cite{Pelissetto:2000ek}). Within Hamiltonian (Lagrangian) approaches there have been several methods proposed to investigate the problem both perturbatively (such as $\varepsilon$ and $1/N$ expansions; see e.g. \cite{Moshe:2003xn}) and non-perturbatively (such as Monte-Carlo simulations), but they may suffer questions of validity and numerical costs.

Recently we have seen substantial progress in understanding higher dimensional CFTs via the conformal bootstrap program \cite{Polyakov:1973ha}\cite{Ferrara:1973yt}. 
%While it only imposes kinematical conditions, 
Without using an explicit form of the Hamiltonian, it provides us with rigorous but non-trivial constraints on spectra of conformal dimensions and operator product expansion (OPE) coefficients \cite{Rattazzi:2008pe}\cite{Caracciolo:2009bx}\cite{Poland:2010wg}\cite{Rattazzi:2010gj}. 
The information encoded in the conformal bootstrap allows us to ``re-discover'' non-trivial CFTs and reproduces known critical exponents \cite{Rychkov:2009ij}\cite{El-Showk:2012ht}\cite{El-Showk:2012hu}\cite{Kos:2013tga}\cite{El-Showk:2013nia}\cite{Beem:2013qxa}\cite{El-Showk:2014dwa}. From our experience, we are convinced that there exist a certain class of CFTs that occupy the extreme corners of the conformal bootstrap constraints although the fundamental reason is not obvious to us yet.

In this Letter we study $O(n)\times O(m)$ symmetric CFTs in space-time dimension $d=3$ with particular emphasis on the {$m=3$} cases. The motivation for this choice of symmetry group is twofold. Firstly, these symmetries are realized in condensed-matter systems like frustrated spin models with non-collinear order \cite{Kawamura}\cite{Delamotte:2003dw} . Secondly, $O(n)\times O(m)$-symmetric Landau-Ginzburg models have richer dynamical structures than $O(n)$ vector models.

There is a long-standing debate whether the frustrated spin systems in non-colinear order show the first order phase transition or the second order phase transition. If the $O(n)\times O(m)$ symmetric Landau-Ginzburg models do not posse non-trivial fixed points other than the $O(nm)$ symmetric Heisernberg fixed point, the second order phase transition is impossible. On the other hand, the existence of the stable $O(n)\times O(m)$ symmetric fixed point suggests the strong evidence for the second order phase transition if the system is inside the attraction domain of the RG flow.

\begin{center}
\begin{figure}[h!!]
  \centering
  \includegraphics[width=8cm]{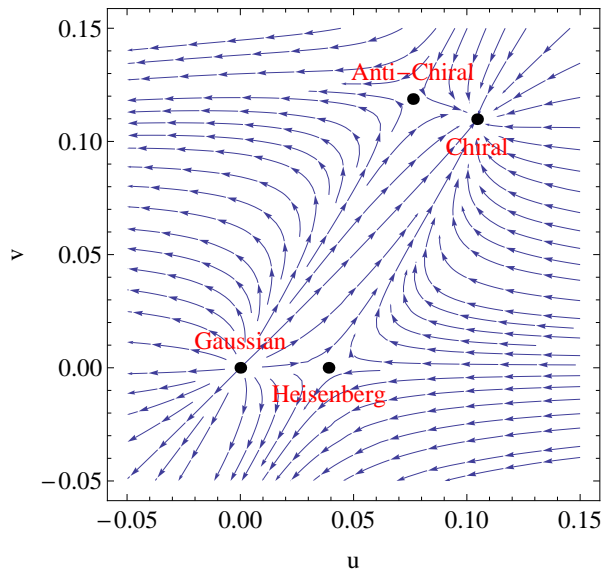}
  \caption{The RG flow of $O(n)\times O(m)$ Landau-Ginzburg models with sufficiently large {$n$}. In the present figure, $(n,m)={(50,3)}$ and the $\beta$-functions are taken from section 3 of \cite{Pelissetto:2001fi}. 
%\green{The chiral fixed point is stable against $(u,v)$ perturbation, while the anti-chiral one has an unstable direction.}
}
  \label{fig:RG}
\end{figure}
\end{center}

When we study the RG flow of the $O(n) \times O(m)$ symmetric Landau-Ginzburg models with the Hamiltonian
\begin{align}
\mathcal{H} &=   \frac{1}{2}(\partial_\mu \phi^\alpha_a)(\partial_\mu \phi^\alpha_a) \cr
&+ \frac{u}{4!} (\phi^{\alpha}_a\phi^\alpha_a)^2 + \frac{v}{4!}(\phi^\alpha_a\phi^\alpha_b \phi^\beta_a \phi^\beta_b  - \phi^\alpha_a\phi^\alpha_a \phi^\beta_b \phi^\beta_b) , \label{Hamiltonian}
\end{align}
where $a=1,\cdots n$ and $\alpha = 1 \cdots m$, we find that for a sufficiently large ratio $n/m$, there are two-more fixed points in addition to Gaussian ($u=v=0$) and $O(nm)$ Heisenberg ($v=0$) fixed points, which are named chiral (stable) and
anti-chiral (unstable) ones \cite{Kawamura}\cite{Pelissetto:2001fi} (see FIG.\ref{fig:RG}). In perturbative approaches, 
the existence of these additional fixed points was established for large $n$ limit with fixed $m$ \cite{Pelissetto:2001fi}, 
but the situations for smaller $n$ was unsettled and controversial \cite{Delamotte:2003dw} \cite{Calabrese:2003ww}\cite{Calabrese:2004nt}, 
so it is desirable to see if we can find any indications of these additional fixed points in the conformal bootstrap program without using the explicit form of the Hamiltonian. In this Letter, we will indeed show strong evidence that for sufficiently large $n/m$, the theories at these fixed points solve the conformal bootstrap constraints at their extreme corners as in \cite{El-Showk:2012ht}\cite{El-Showk:2014dwa} but in different OPE sectors. Furthermore we will predict the edge of the conformal window for ${O(n)\times O(3)}$ anti-chiral fixed points.

Our careful choice of $m=3$ is to establish the usefulness of the conformal bootstrap program to determine the edge of the  conformal window  in less controversial situations \cite{Parruccini} while still finding the conjecture from the other methods before we venture into the the most controversial cases of $m=2$ with smaller $n$. Note that $m=3$ cases are not necessarily unrealistic (compared e.g. with $m>3$). As reviewed in \cite{Kawamura}, when the frustrated spins align three-dimensionally, the system admits the $O(n) \times O(3)$ symmetry.

\section{Conformal Bootstrap}
Let us begin with CFTs with no continuous global symmetry. The strategy of the conformal bootstrap program is to constrain CFTs from their %kinematical 
fundamental properties such as conformal invariance, unitarity and crossing symmetry. Specifically, we demand these properties in the four-point function of identical scalar primary operators,
\begin{align*}
 \langle \phi (x_1 ) \phi (x_2) \phi (x_3 ) \phi (x_4) \rangle.
\end{align*}
The crossing relation gives a sum rule in terms of OPE coefficients and conformal blocks. After truncation to a finite dimensional convex optimization problem, we find an upper bound for the conformal dimension of the lowest-dimensional operator (other than the unit operator) of definite spin appearing in $\phi \times \phi$ OPE. Repeating this procedure by changing the conformal dimension $\delta$ of the external scalar operator, we obtain a critical curve $\Delta _c ^l (\delta)$ with $l$ being the spin of the operator. The behavior of $\Delta_c^0 (\delta)$ is summarized as: for space-time dimensions $d <4$, it shows a discontinuity in its slope, and the location of such a ``kink'' corresponds to the Wilson-Fisher fixed point. The singular behavior of $\Delta_c^0 (\delta)$ knows the existence of a non-trivial CFT.

Let us now consider the CFT with an additional symmetry group $G$.
We study the four-point function of scalar operators in a certain real representation $R$ of $G$, 
\begin{align*} 
 \langle \phi_i (x_1 ) \phi_j (x_2) \phi_k (x_3 ) \phi_l (x_4) \rangle, 
\end{align*}
where $i$ is the index for the representation $R$. In these cases, intermediate states are labelled by their spin and the representation appearing in $R\otimes R$, and the crossing relation gives a vectorial sum rule \cite{Rattazzi:2010yc}\cite{Vichi:2011ux}\cite{Poland:2011ey}\cite{Kos:2013tga}. Correspondingly, we find a bound for the conformal dimension of the lowest operator appearing in $\phi _i \times \phi _j$ OPE for each irreducible representation contained in $R\otimes R$ to obtain a critical curve $\Delta_c ^{I,l}(\delta)$ ($I$ and $l$ denoting the representation and spin). Note that the possible representation $I$ and parity of $l$ are correlated by Bose symmetry of the operator.

In \cite{Kos:2013tga}, they studied the case of $G=O(n)$ in $d=3$ with $R$ given by fundamental representation in detail. In this case we have three irreducible representations, singlet (which we denote $\mathrm{S}$), second-rank symmetric tensor ($\mathrm{T}$), and anti-symmetric tensor representation ($\mathrm{A}$). In the $\mathrm{S}$ sector, they found kinks (though the changes of the slope were milder) in $\Delta_c ^{\mathrm{S},0}(\delta)$, the location of which exhibits an excellent agreement with that of $O(n)$ Heisenberg fixed points proposed in the other methods (e.g. large $n$ expansion). They also obtained the bounds for scalar operators in the $\mathrm{T}$ sector. The resulting $\Delta_c ^{\mathrm{T},0} (\delta)$ reveals intriguing but not as prominent features: at the point where the $O(n)$ model sits, it starts to grow approximately linearly.

Our focus is $ G = O(n) \times O(m)$ (with $n, m \ge 3$ for simplicity) in $d=3$ under the presence of a scalar operator in bifundamental representation $R$ corresponding to $\phi^{\alpha}_a$ in \eqref{Hamiltonian}. In this case, $R \otimes R$ contains nine irreducible representations, which are product representations formed by $\mathrm{S}$, $\mathrm{T}$, and $\mathrm{A}$. The sum rule encoding this information is conveniently expressed in terms of a $9 \times 9$ matrix, which we have derived following the line of \cite{Poland:2011ey}.

To compute numerical bounds, we follow the methods described in \cite{Kos:2013tga} though the details are somewhat distinct. For example, we include intermediate operators with spin $l \le 20$. Since the sum rule matrix is larger and computational task is much heavier than $O(n)$ case, we work in lower dimensional search spaces, namely $36 \times 9$-dimensional ones (or $k=8$ in \cite{Kos:2013tga}, $N_{\mathrm{max}}=7$ in \cite{El-Showk:2012ht}). In \cite{Kos:2013tga}, they obtained the bounds by assuming the conformal dimension of intermediate scalar operator in the $\mathrm{T}$ sector to be greater than 1 for technical reasons, but we only impose the unitarity bound in every sectors. We use the $\rho$ - series expansions in \cite{Hogervorst:2013kva} to generate the residues of the conformal blocks. Our normalization condition of the linear functional $\Lambda$ is such that it gives the value $1$ when it acts on the vector for a dimension $5$ conformal block in the spin 0 $\mathrm{TT}$ sector. Our implementation for \verb+sdpa-gmp+ \cite{sdpa1}\cite{sdpa2} is the same as in \cite{Kos:2013tga} except that the parameter \verb+precision+ is $350$.

\section{Results}
We first performed the numerical computation for ${O(15) \times O(3)}$ model since the value ${n}=15$ is well above the existence bound on additional chiral/anti-chiral fixed points obtained from the large ${n}$ analysis of \cite{Pelissetto:2001fi}. 

\begin{center}
\begin{figure}[h!!!]
  \centering
  \includegraphics[width=8.5cm]{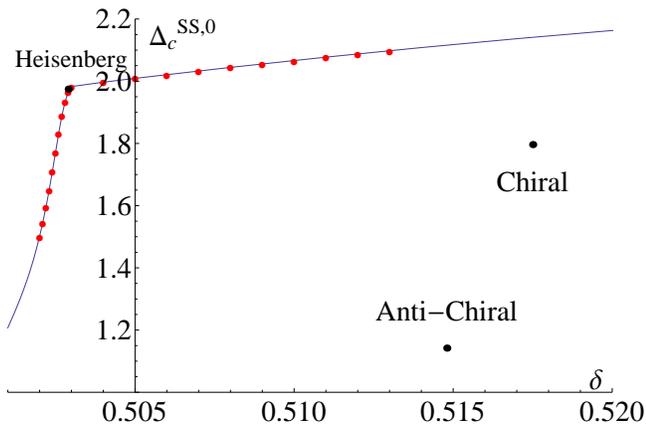}
  \caption{The plot for $\Delta_c^{\mathrm{SS},0}(\delta)$ for $O(15)\times O(3)$(red dots) and $\Delta _c ^{\mathrm{S},0}$ for $O(45)$ (blue line): The vertical precision is $10^{-3}$. The location of Heisenberg fixed point is from the large $nm$ analysis reviewed in \cite{Moshe:2003xn}. Those of chiral/anti-chiral fixed points are the large $n$ predictions of \cite{Pelissetto:2001fi}. }
  \label{fig:1}
\end{figure} 
\end{center}

\begin{center}
\begin{figure}[h!!!]
  \centering
  \includegraphics[width=7cm]{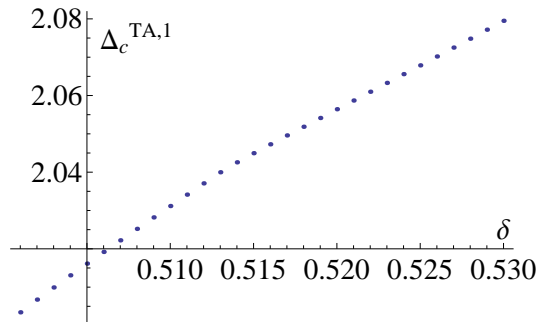}
  \caption{The plot for $\Delta_c^{{\mathrm{TA}},1}(\delta)$: The vertical precision is $4\times 10^{-4}$. See also FIG. 7 for the differentiated plot.}
  \label{fig:2}
\end{figure}
\end{center}

{\bf Symmetry enhancement in the singlet sector.} We computed the bounds for the first scalar operator in the $\mathrm{SS}$ sector, $\Delta_c^{\mathrm{SS},0} (\delta )$. We present the results in FIG. \ref{fig:1}. They turned out to be identical within the precision of $10^{-4}$ to those of the $\mathrm{S}$ sector in $O(45)$ model. Such ``symmetry enhancement'' behavior was reported in \cite{Poland:2011ey} between $SU(N)$ and $O(2N)$. As stressed there, we can prove the inequality $\Delta_{c,O(45)} ^{\mathrm{S},0}(\delta) \le \Delta _{c,{O(15)\times O(3)}}^{\mathrm{SS},0} (\delta)$ immediately because we can regard every CFT with $O(45)$ symmetry as a CFT with ${O(15) \times O(3)}$ symmetry upon decomposing $O(45)$ representations into those of ${O(15)\times O(3)}$. We therefore conclude that the kink shown in the SS sector corresponds to the Heisenberg fixed point. 
However, the reason for the actual equality here is still mysterious. Meanwhile, chiral/anti-chiral fixed points lie well below the bound, which confirms the consistency of our analysis, but simultaneously implies that we will not be able to approach these fixed points in this manner. We propose an alternative way to spot them in the rest of the present section.
{\bf The bound and spectral study for the {{TA}} spin $1$ operator.} We next computed the bounds for spin 1 operators in {$O(15)$ symmetric, $O(3)$ anti-symmetric} tensor representation, $\Delta_c ^{{\mathrm{TA}},1}(\delta)$. Such an operator would be conserved if {$O(15) \times O(3)$} were enhanced to $O(45)$. The result in FIG. \ref{fig:2} shows a kink behavior around $\delta \sim 0.515(1)$. This value is quite close to the value $\delta = 0.5148$ predicted by the large {$n$} analysis \cite{Pelissetto:2001fi} of the anti-chiral fixed point (c.f. $\delta = 1/2 + \eta/2$). For a further check of this identification, we have derived {low-lying} spectra at the kink from the \verb+sdpa-gmp+ output, following the strategy of \cite{El-Showk:2012hu}. The first operator in the $\mathrm{SS}$ sector has the conformal dimension $1.16(3)$. On the other hand, the large {$n$} analysis predicts that it is $1.142$ in close agreement with ours (c.f. $\Delta^{\mathrm{SS,0}} = 3-1/\nu$). We estimate the systematic errors conservatively from the vertical bisection precision, the horizontal impreciseness to locate the kink as well as the convergence with respect to the number of derivatives in our search space.
 For reference, we also computed the low-lying spectra in the other sectors from the same output. For the $\mathrm{SA}$ sector the first operator has the conformal dimension $2.02$ while it should be exactly $2$ since it must be a conserved current. Hence the error in this analysis could be as large as 0.02, and our prediction for the $\mathrm{SS}$ scalar does not seem to contradict with the large {$n$} analysis.

{\bf Other sectors.} We also computed the numerical bounds for the other sectors with the lowest spins.

\begin{center}
\begin{figure}[h!!!]
  \centering
  \includegraphics[width=7cm]{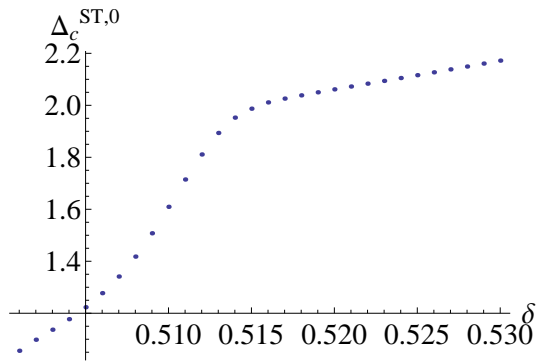}
  \caption{The plot for $\Delta_c^{{\mathrm{ST}},0}(\delta)$: The vertical precision is $4 \times 10^{-4}$.}
  \label{fig:3}
\end{figure}
\end{center}

For the {$\mathrm{ST}$} spin 0 sector, we present the results in FIG. \ref{fig:3}. While the change of the slope is not as sharp as that in the {$\mathrm{TA}$} sector, the shape resembles the bound for {the spin 0 operator} in the 2d Ising model reported in \cite{Rychkov:2009ij} with $k=8$ (more recently the bound has been improved to give a sharper kink behavior, see \cite{El-Showk:2012hu}). It might be feasible to sharpen the bound so that this becomes an actual kink. The spectra read off at $\delta = 0.515$ shows that the conformal dimension of the $\mathrm{SS}$ operator is $1.16(3)$, which is again close to the large {$n$} prediction for the anti-chiral fixed point.

\begin{center}
\begin{figure}[h!!!]
  \centering
  \includegraphics[width=7cm]{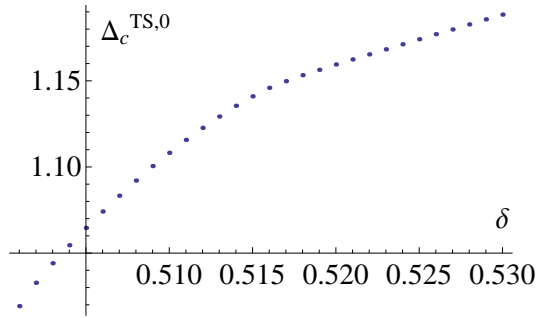}
  \caption{The plot for $\Delta_c^{{\mathrm{TS}},0}(\delta)$: The vertical precision is $2\times 10^{-4}$.}
  \label{fig:4}
\end{figure}
\end{center}

For the ${\mathrm{TS}}$ spin 0 sector, we present the results in FIG. \ref{fig:4}. We see no sudden change of the slope, but its behavior looks similar to the $\mathrm{T}$ sector bounds in \cite{Kos:2013tga}, so we conjecture that there is a nontrivial CFT saturating the bound at the point where $\Delta _c ^{{\mathrm{TS}},0}(\delta)$ starts to behave linearly, that is, at $\delta = 0.517(1)$. Reading off the spectra at $\delta = 0.517$, we obtained the conformal dimension of the $\mathrm{SS}$ scalar as $1.81(3)$. The large {$n$} analysis at the chiral fixed point predicts that $\delta$ is $0.5175$ and the conformal dimension of the first $\mathrm{SS}$ scalar is $1.796$, which is once again close to the present estimate of ours.

On the other hand, $\mathrm{TT}$ and $\mathrm{AA}$ spin $0$ sectors indicate a weakly kink-like behavior near the Heisenberg fixed point with similar spectra, but {$\mathrm{AT}$} spin $1$ sector shows no interesting behavior at all. It is not clear to us why such a preference among different sectors exists.

\begin{center}
\begin{figure}[h!!!]
  \centering
  \includegraphics[width=9cm]{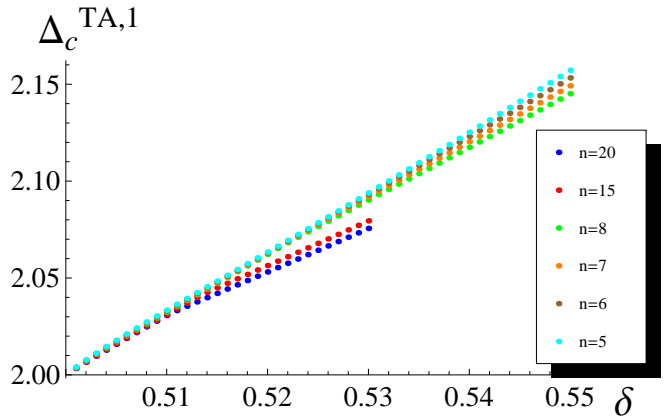}
  \caption{The plot for $\Delta_c ^{{\mathrm{TA}},1} (\delta)$ with ${n}=20,8,7,6,5$: The vertical precisions with {$n=20, 15, 8$ } are $10^{-4}$, while those with $n=7,6,5$ are $2\times 10^{-5}$}
  \label{fig:5}
\end{figure}
\end{center}

\begin{center}
\begin{figure}[h!!!]
  \centering
  \includegraphics[width=8cm]{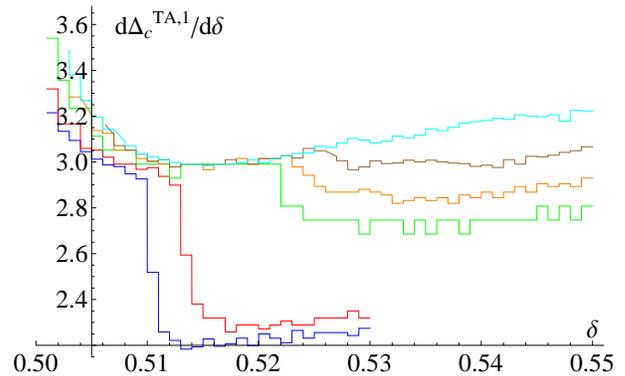}
  \caption{The slopes of FIG. \ref{fig:5}: The legends are same as in FIG. \ref{fig:5}. %For {$n=20$} and $\delta >0.53$, the slope is extrapolated.%
}
  \label{fig:6}
\end{figure}
\end{center}

{\bf Prediction of conformal window.} Now that we have demonstrated that anti-chiral fixed points show up in the {$\mathrm{TA}$} spin 1 operator bound for {$n=15$}, we want to apply the same method to determine the edge of the conformal window under which the anti-chiral fixed point disappears. To do this we computed {$\mathrm{TA}$} bounds for {$n = 20, 8,7,6,5$}. We present the results in FIG. \ref{fig:5}. For convenience we complement the plot of its first derivatives (generated by \verb+Interpolation+ function of \verb+Mathematica+) in FIG. \ref{fig:6}. As we decrease {$n$}, the change of the slope decreases and disappears at {$n=6 \sim 7$}. Thus we predict that the edge of the conformal window for anti-chiral fixed points is $7\sim 8$. In comparison, we quote that the predicted value from the large {$n$} analysis in \cite{Pelissetto:2001fi} was $7.3$. See also TABLE I of \cite{Calabrese:2003ww} for the summary of the predictions from the other approaches.
We observed there is a wider discrepancy between the location of the kink and the large {$n$} prediction of $\delta$ for anti-chiral fixed points toward smaller {$n$} around $8$. It is interesting to see if the analysis with higher dimensional search space would resolve this gap.
\section{Concluding Remarks}
In this Letter, we have accomplished the conformal bootstrap program for {$O(n)\times O(3)$} symmetric CFTs with various {$n$}. In particular, the detailed study for {$n=15$} has revealed the existence of singular behaviors in numerical bounds corresponding to Heisenberg as well as chiral and anti-chiral fixed points of the RG flow, depending on the sector for which we computed the bound. Moreover we have predicted the edge of the conformal window for anti-chiral fixed points. We emphasize that not only can our results serve as a novel tool for the study of critical phenomena, but they are encouraging for the conformal bootstrap program itself because our result makes it evident that {\it different} kind of CFTs are hidden behind a {\it single} (vectorial) bootstrap equation.

There are several future directions to pursue. The obvious one is to study other symmetry groups. Among them, {$O(n) \times O(2)$} will be particularly important in the context of condensed-matter physics. For QCD applications in mind, the similar analysis for $U(n) \times U(m)$ groups is under study \cite{appear}. We could also refine our results by extending the search space to include larger number of derivatives. This might fill the observed gap between our analysis and the large {$n$} one. Our prediction of the conformal window is based on the anti-chiral fixed point. If the RG picture that the disappearance of the anti-chiral fixed point is induced by the annihilation with the chiral fixed point, the same conformal window should have been obtained from the chiral fixed point seen in ${\mathrm{TS}}$ spin 0 sector by changing $n$. This may not be the case as proposed in \cite{Calabrese:2004nt} for $m=2$ cases, and it should be interesting to see it directly in our approach. Since the signal is weaker, however, it may require the wider search spaces of conformal bootstrap program with more CPU power needed.

Finally, it has been recently conjectured that the 3d Ising model can be characterized as the CFT which minimizes the central charge in the space of entire CFTs \cite{El-Showk:2014dwa}. In analogy, we speculate the possibility if we can characterize the CFTs at chiral/anti-chiral fixed points as those extremizing some quantities (e.g. current central charges). 
\section*{Acknowledgements}
We would like to thank P.~Calabrese, S.~Rychkov, D.~Simmons-Duffin and E.~Vicari for correspondence and discussions.
This work is supported by the World Premier International Research Center Initiative (WPI Initiative), MEXT. T.O. is supported by JSPS Research Fellowships for Young Scientists and the Program for Leading Graduate Schools, MEXT.
\bibliographystyle{ytphys}
%\bibliography{LGref}

\end{document}